\documentclass[a4paper,twocolumn,showpacs,nofootinbib,floatfix,superscriptaddress,prc]{revtex4}
\usepackage{graphicx}
\usepackage{amsfonts}
\usepackage{amsbsy}
\usepackage{longtable}
\usepackage{bm}

\renewcommand{\vec}[1]{\mathbf{#1}}
\def\be{\begin{equation}}
\def\ee{\end{equation}}
\def\bea{\begin{eqnarray}}
\def\eea{\end{eqnarray}}
\def\simge{\mathrel{
    \rlap{\raise 0.511ex \hbox{$>$}}{\lower 0.511ex \hbox{$\sim$}}}}
\def\simle{\mathrel{
    \rlap{\raise 0.511ex \hbox{$<$}}{\lower 0.511ex \hbox{$\sim$}}}}

\begin{document}
%%%%%%%%%%%%%%%%%%%%%%%%%%%%%%%%%%%%%%%%%%%%%%%%%%%%%%%%%%%

\title{Entropy production in high-energy heavy-ion collisions and the
  correlation of shear viscosity and thermalization time}

\bigskip

\author{A.~Dumitru}
\affiliation{Institut f\"ur Theoretische Physik, Johann Wolfgang Goethe Universit\"at,\\
Max-von-Laue-Str. 1, 60438 Frankfurt am Main, Germany}

\author{E.~Moln\'ar}
\affiliation{Frankfurt Institute for Advanced Studies, Johann Wolfgang Goethe
Universit\"at, Max-von-Laue-Str. 1, 60438 Frankfurt am Main, Germany }

\author{Y.~Nara}
\affiliation{Institut f\"ur Theoretische Physik, Johann Wolfgang Goethe Universit\"at,\\
Max-von-Laue-Str. 1, 60438 Frankfurt am Main, Germany}
\affiliation{Akita International University 193-2 Okutsubakidai, Yuwa-Tsubakigawa, Akita 010-1211, Japan}

%\date{\today}

\bigskip

\begin{abstract}
We study entropy production in the early stage of high-energy
heavy-ion collisions due to shear viscosity. We employ the
second-order theory of Israel-Stewart with two different stress
relaxation times, as appropriate for strong coupling or for a
Boltzmann gas, respectively, and compare the hydrodynamic evolution.
Based on present knowledge of initial particle production, we argue
that entropy production is tightly constrained. We derive new limits
on the shear viscosity to entropy density ratio $\eta/s$, independent
from elliptic flow effects, and determine the corresponding Reynolds
number. Furthermore, we show that for a given entropy production
bound, that the initial time $\tau_0$ for hydrodynamics is correlated
to the viscosity. The conjectured lower bound for $\eta/s$ provides a
lower limit for $\tau_0$.
\end{abstract}


\maketitle

%%%%%%%%%%%%%%%%%%%%%%%%%%%%%%%%%%%%%%%%%%%%%%%%%%%%%%%%%%%%%%%%%%%%%%%%%%%
\section{Introduction}
%%%%%%%%%%%%%%%%%%%%%%%%%%%%%%%%%%%%%%%%%%%%%%%%%%%%%%%%%%%%%%%%%%%%%%%%%%%

Experiments with colliding beams of gold ions at the Relativistic
Heavy-Ion Collider (RHIC) have confirmed that dense QCD matter
exhibits hydrodynamic flow effects~\cite{flowReview}. Their magnitude
matches approximately predictions based on ideal Euler (inviscid)
hydrodynamics\footnote{Numerical solutions of Euler hydrodynamics on
finite grids always involve some amount of numerical viscosity for
stability. Reliable algorithms such as flux-corrected transport keep
this numerical viscosity and the associated entropy production at a
minimum~\cite{NumVisco}.}~\cite{Huovinen:2006jp}. More precisely, the
transverse momentum and centrality dependence of the azimuthally
asymmetric flow, $v_2$, requires a shear viscosity to entropy density
ratio as low as $\eta/s\le0.2$~\cite{Teaney,Lacey:2006bc,DDGO,RR07};
this is much lower than perturbative extrapolations to temperatures
$T\simeq200$~MeV~\cite{Csernai_1}. However, it is comparable to recent
results for SU(3) pure gauge theory from the lattice~\cite{LatVisco},
and to the conjectured lower bound for strongly coupled systems
$\eta/s\ge1/(4\pi)$~\cite{Kovtun_1}. Similar constraints on $\eta/s$
have been derived from transverse momentum
correlations~\cite{GavinAziz} and from energy loss and flow of heavy
quarks at RHIC~\cite{phenixQ}.

The purpose of this paper is to obtain an independent upper bound on
$\eta/s$ by analyzing entropy production in the early stages of the
hydrodynamic evolution (the plasma phase), where the expansion rate
and hence the entropy production rate is largest. Entropy production
in heavy-ion collisions due to viscous effects has been studied
before~\cite{EntroHIC,Muronga}. The new idea pursued here is that
recent progress in our understanding of gluon production in the {\em
initial} state constrains the amount of additional entropy produced
via ``final-state'' interactions, and hence the viscosity and the
thermalization time. The second-order formalism for viscous
hydrodynamics of Israel and Stewart~\cite{Israel_Stewart}, and its
application to one-dimensional boost-invariant Bjorken
expansion~\cite{Bjorken:1982qr}, are briefly reviewed in
section~\ref{DissHydro}.

The initial condition for hydrodynamics, in particular the initial
parton or entropy density in the central rapidity region, plays a
crucial role. If it is close to the measured final-state multiplicity,
this provides a stringent bound on viscous effects. The initial parton
multiplicity in heavy-ion collisions can, of course, not be measured
directly. Our analysis therefore necessarily relies on a calculation
of the initial conditions (presented in
section~\ref{CGC_IC}). Specifically, we employ here a
$k_\perp$-factorized form of the ``Color Glass Condensate'' (CGC)
approach which includes perturbative gluon saturation at small
light-cone momentum fractions $x$~\cite{KLN}. However, different
approaches for initial particle production, such as the HIJING model
which relies on collinear factorization supplemented with an
additional model for the soft regime, also predicts multiplicities
close to experiment~\cite{hijing}. The same is true when the heavy-ion
collision is modeled as a collision of two classical Yang-Mills
fields~\cite{KNV}. It is important to test these models for small
systems, such as peripheral $A+A$ (or even $p+p$) collisions, in order
to constrain the entropy increase via final-state effects
(thermalization and viscosity).

Section~\ref{Results} contains our main results. We show how the
entropy production bound correlates $\eta/s$ to the initial time for
hydrodynamic evolution, $\tau_0$. The entropy production rate grows
with the expansion rate (i.e., how rapidly flow lines diverge from
each other), and the total amount of produced entropy is therefore
rather sensitive to the early stages of the expansion. The bound on
the viscosity depends also on the initial condition for the stress,
which in the second-order theory is an independent variable and is not
fixed by the viscosity and the shear (unless the stress relaxation
time is extremely short, as predicted recently from the AdS/CFT
correspondence at strong coupling~\cite{Heller}).

In a recent paper, Lublinsky and Shuryak point out that if the initial
time $\tau_0$ is assumed to be very small, that a resummation of the
viscous corrections to all orders in gradients of the velocity field
is required~\cite{LS07}. Here, we explore only the regime where
$\tau_0$ is several times larger than the sound attenuation length
$\Gamma_s$, and so the standard approach to viscous hydrodynamics
should apply. Quantitatively, we find that an entropy bound of
$\simeq10\%$ restricts $\eta/s$ to be at most a {\em few} times the
lower bound ($\eta/s=1/(4\pi)$) conjectured from the
correspondence~\cite{Kovtun_1}. On the other hand, somewhat
surprisingly, we find that even $\eta/s=1/(4\pi)$ is large enough to
give noticeable entropy production for thermalization times $\tau_0$
well below 1~fm/c (but still larger than $\Gamma_s$). Present
constraints from initial and final multiplicities are not easily
reconciled with such extremely short initial
times~\cite{Kovchegov:2007pq}.

We restrict ourselves here to 1+1D Bjorken expansion. Given its
(numerical) simplicity and the fact that the entropy production rate
is largest at early times, this should provide a reasonable starting
point. Estimates for the initial time $\tau_0$, for the parton density
and the stress at $\tau_0$, and for the viscosity to entropy density
ratio $\eta/s$ are in fact most welcome for large-scale numerical
studies of relativistic (second-order) dissipative fluid
dynamics. Without guidance on the initial conditions, the hydrodynamic
theory can at best provide qualitative results for heavy-ion
collisions.

We neglect any other possible source of entropy but shear viscosity at
early times\footnote{This contribution is expected to vanish once transverse
expansion is fully developed, see sect.~\ref{Diss_Bjorken}.}. Even
within this simplified setting, there could be additional entropy
production due to a viscous ``hadronic corona'' surrounding the
fireball~\cite{HiranoGy}, which we do not account for. Clearly, any
additional contribution would further tighten the (upper) bound on
$\eta/s$ and the (lower) bound on $\tau_0$.  We also assume that
$\eta/s$ is constant. This does not hold over a very broad range of
temperature~\cite{Csernai_1} but should be a reasonable first
approximation for $T\simeq 200$-400~MeV.

We employ natural units throughout the paper: $\hbar=c=k_B=1$.

%%%%%%%%%%%%%%%%%%%%%%%%%%%%%%%%%%%%%%%%%%%%%%%%%%%%%%%%%%%%%%%%%%%%%%%%%%%%%%%
\section{Dissipative fluid dynamics}  \label{DissHydro}
%%%%%%%%%%%%%%%%%%%%%%%%%%%%%%%%%%%%%%%%%%%%%%%%%%%%%%%%%%%%%%%%%%%%%%%%%%%%%%%

%%%%%%%%%%%%%%%%%%%%%%%%%%%%%%%%%%%%%%%%%%%%%%%%%%%%%%%%%%%%%%%%%%%%%%%%%%%%%%%
\subsection{Second-order formalism}
%%%%%%%%%%%%%%%%%%%%%%%%%%%%%%%%%%%%%%%%%%%%%%%%%%%%%%%%%%%%%%%%%%%%%%%%%%%%%%%

In this section we briefly review some general expressions for viscous
hydrodynamics which will be useful in the following. More extensive
discussions are given in
refs.~\cite{Muronga,Heinz,MuroRi,BRW,Romatschke,Romatschke2,Muronga2,Bhalerao:2007ek},
for example.

A single-component fluid is generally characterized by a conserved current
(possibly more), $N^{\mu}$, the energy-momentum tensor $T^{\mu \nu}$
and the entropy current $S^{\mu}$. The conserved quantities satisfy
continuity equations,
\bea
\partial_{\mu} N^{\mu} = 0 \, , \qquad \partial_{\mu} T^{\mu \nu} = 0\, .
\eea
In addition, the divergence of the entropy current has to be positive
by the second law of thermodynamics,
\be\label{entropy_1}
\partial_{\mu} S^{\mu} \geq 0 \, .
\ee
For a perfect fluid, a well-defined initial-value problem requires the
knowledge of $T^{\mu\nu}$ and of $N^\mu$ on a space-like surface in
3+1D Minkowski space-time. This is equivalent to specifying the
initial flow field $u^{\mu}$, the proper charge density $n\equiv u_\mu
N^\mu$, and the proper energy density $e\equiv u_\mu u_\nu
T^{\mu\nu}$; the pressure is determined via an algebraic relation to
$e$ and $n$, the equation of state (EoS).

In dissipative fluids, irreversible viscous and heat conduction
processes occur.  These quantities can be expressed explicitly if the
charge and entropy currents and the energy-momentum tensor are
decomposed (projected) into their components parallel and
perpendicular to the flow of matter~\cite{Eckart}; the latter describe
the dissipative currents. The transverse projector is given by
$\Delta^{\mu \nu}= g^{\mu \nu} - u^\mu u^\nu$, with $g^{\mu \nu} =
\textrm{diag} (1,-1,-1,-1)$ the metric of flat space-time.
In the following, we focus on locally charge-neutral systems where all
conserved currents vanish identically.

The energy-momentum tensor can be decomposed in the following way:
\be T^{\mu \nu}
= e u^{\mu} u^{\nu} - (p+\Pi) \Delta^{\mu \nu} + W^{\mu} u^{\nu} + W^{\nu}
u^{\mu} + \pi^{\mu \nu}~. \label{EMtensor1}
\ee
Here, $W^{\mu} = q^{\mu} + h V^{\mu}= u_{\nu} T^{\nu \alpha}
\Delta^{\mu}_{\alpha}$ is the energy flow, with $h = (e + p)/n$ the
enthalpy per particle, and $q^{\mu}$ is the heat flow; we shall {\em
define} the local rest-frame via $W^\mu=0$ (the ``Landau frame'').
Furthermore, $\Pi$ denotes the bulk pressure such that $p + \Pi = -\frac{1}{3}
\Delta_{\mu \nu} T^{\mu \nu}$, while the symmetric and traceless part of the
energy-momentum tensor defines the stress tensor, $\pi^{\mu \nu} =
\left[\frac{1}{2} \left( \Delta^{\mu}_{\alpha} \Delta^{\nu}_{\beta} +
\Delta^{\nu}_{\alpha} \Delta^{\mu}_{\beta} \right) - \frac{1}{3}
\Delta^{\mu \nu} \Delta_{\alpha \beta} \right] T^{\alpha \beta}$.

The entropy current is decomposed as
\be
S^{\mu} = s u^{\mu} + \Phi^{\mu}~.
\ee
In the standard first order theory due to Eckart~\cite{Eckart} and
Landau and Lifshitz~\cite{Landau_book}, only linear corrections are
taken into account, i.e., $\Phi^{\mu} = q^{\mu}/T$.  On the other
hand, the second order theory of relativistic dissipative fluid
dynamics includes terms to second order in the irreversible flows and
in the stress tensor~\cite{Israel_Stewart}:
\bea \label{entropy_2}
S^{\mu} &=&
  s u^{\mu} + \frac{q^{\mu}}{T} - \left( \beta_0 \Pi^2 - \beta_1 q_{\nu}
  q^{\nu} + \beta_2 \pi_{\nu \alpha} \pi^{\nu \alpha}\right)
  \frac{u^{\mu}}{2T} \nonumber\\
& & - \frac{\alpha_0 \Pi q^{\mu}}{T} +
\frac{\alpha_1 \pi^{\mu \nu} q_{\nu}}{T} ~,
\eea
where the coefficients $\beta_0$, $\beta_1$, $\beta_2$ and $\alpha_0$,
$\alpha_1$ represent thermodynamic integrals which (near equilibrium)
are related to the relaxation times of the dissipative
corrections. Furthermore, from~(\ref{entropy_1}), one can find linear
relationships between the thermodynamic forces and fluxes, leading to
the transport equations describing the evolution of dissipative
flows~\cite{Israel_Stewart}.

In what follows, we will focus on shear effects and neglect heat flow
and bulk viscosity, hence~(\ref{EMtensor1}) simplifies to
\be
T^{\mu \nu}
= e u^{\mu} u^{\nu} - p\Delta^{\mu \nu} + \pi^{\mu \nu}~. \label{EMtensor2}
\ee
The stress tensor satisfies a relaxation equation,
\be \label{pi_sigma_IS}
\tau_\pi u^\lambda\partial_\lambda \pi^{\mu\nu} + \pi^{\mu\nu} =
  2\eta \sigma^{\mu\nu}~,
\ee
where $\eta$ denotes the shear viscosity, and the shear tensor
$\sigma^{\mu\nu}$ is a purely ``geometrical'' quantity, determined by
the flow field:
\be
\sigma^{\mu\nu} = \frac{1}{2} \left( \nabla^\mu u^\nu + \nabla^\nu
                                 u^\mu\right)
  - \frac{1}{3} \Delta^{\mu\nu} \nabla_\lambda u^\lambda ~,
\ee
with $\nabla^\mu=\Delta^{\mu\nu} \partial_\nu$. The relaxation time
$\tau_\pi$ determines how rapidly the stress tensor $\pi^{\mu\nu}$
relaxes to the shear tensor $\sigma^{\mu\nu}$; in particular, in the
limit $\tau_\pi\to0$,
\be   \label{pi_sigma_NS}
\pi^{\mu\nu}=2\eta \sigma^{\mu\nu}
\ee
satisfy the same algebraic relation as in the first-order theory.  The
limit $\tau_\pi\to0$ is formal, however, since the deviation of the
stress $\pi^{\mu\nu}$ from $2\eta\sigma^{\mu\nu}$ at any given time,
as obtained by solving eq.~(\ref{pi_sigma_IS}), depends also on its
initial value. If~(\ref{pi_sigma_NS}) is approximately valid
at the initial time then the first-order theory may provide a
reasonable approximation for the entire evolution (see below).

By analyzing the correlation functions of the stress that lead to the
definitions~(\ref{pi_sigma_IS},\ref{pi_sigma_NS}), respectively, of
the shear viscosity, Koide argues that in the second-order theory of
Israel and Stewart $\eta$ may represent a different quantity than in
the first-order approach~\cite{Koide}. Nevertheless, here we assume
that the conjectured lower bound for $\eta/s$ applies even to the
causal (second-order) approach.

%%%%%%%%%%%%%%%%%%%%%%%%%%%%%%%%%%%%%%%%%%%%%%%%%%%%%%%%%%%%%%%%%%%%%%%%%%%%%%
\subsection{Dissipative Bjorken scaling fluid dynamics} \label{Diss_Bjorken}
%%%%%%%%%%%%%%%%%%%%%%%%%%%%%%%%%%%%%%%%%%%%%%%%%%%%%%%%%%%%%%%%%%%%%%%%%%%%%%%

In this section we recall the 1+1D Bjorken scaling
solution~\cite{Bjorken:1982qr} in 3+1D space-time including
stress~\cite{Muronga}. By assumption, the fluid in the central region
of a heavy-ion collision expands along the longitudinal $z$-direction
only, with a flow velocity $v$ equal to $z/t$. This is appropriate for
times less than the transverse diameter $R$ of the collision zone
divided by the speed of sound $c_s=\sqrt{\partial p/\partial e}$
(possibly longer for very viscous fluids). After that transverse
expansion is fully developed and we expect that entropy production due
to shear decreases. In fact, it is straightforward to check that for
{\em three-dimensional} scaling flow\footnote{Replace $z\to|\bm{r}|$
in the definition of proper time $\tau$ and space-time rapidity $\tilde\eta$
below, and take $\bm{u}=\bm{r}/\tau$.} $u_\mu \partial_\nu
\sigma^{\mu\nu}=0$; hence, within the first-order theory at least, the
shear viscosity does not enter the evolution equation of the energy
density anymore.

Formulations of the Israel-Stewart second-order theory for Bjorken
plus transverse expansion have been
published~\cite{Heinz,MuroRi,BRW,Romatschke,Romatschke2,Muronga2} but require
large-scale numerical computations. A relatively straightforward 1+1D
analysis is warranted as a first step to provide an estimate for
entropy production.

It is convenient to transform from $(t,z)$ to new $(\tau,\tilde\eta)$
coordinates, where $\tau = \sqrt{t^2-z^2}$ denotes proper time and
$\tilde\eta = \frac{1}{2} \log ((t+z)/(t-z))$ is the space-time rapidity;
for the Bjorken model, it is equal to the rapidity of the flow,
$\eta_{\rm fl} \equiv \frac{1}{2} \log ((1+v)/(1-v))$. In other words,
the four-velocity of the fluid is $u^\mu = (\delta^{\mu 0} +
\delta^{\mu 3})\, x^\mu/\tau$.

The longitudinal projection of the continuity equation for the
stress-energy tensor then yields
\be
\frac{d e}{d\tau} + \frac{e + p}{\tau} - \frac{\Phi}{\tau} +
\frac{\Pi}{\tau} = 0\,.   \label{eq:de_dtau}
\ee
Here, $e$ is the energy density of the fluid in the local rest-frame,
while $p$ denotes the pressure. These quantities are related through
the equation of state (EoS). We focus here on entropy production
during the early stages of the evolution where the temperature is
larger than the QCD cross-over temperature $T_c\simeq 170$~MeV, and so
assume a simple ideal-gas EoS, $p=e/3$.

In what follows, we will neglect the bulk pressure $\Pi$ which would
otherwise tend to increase entropy production further. Well above
$T_c$, this contribution is expected to be much smaller than that due
to shear~\cite{Arnold:2006fz}. In the transition region, the bulk
viscosity could be significant~\cite{bulk}.  Also, for 1+1D expansion
considered here, the stress $\Phi\equiv \pi^{00} - \pi^{zz}$
acts in the same way as the bulk pressure $\Pi$:
only the combination $\Phi-\Pi$ appears in~(\ref{eq:de_dtau}).

The time evolution of the stress is determined by~\cite{Muronga}
\be\label{phi}
\frac{d\Phi}{d\tau} + \frac{\Phi}{\tau_{\pi}} + \frac{\Phi}{2}
\left[ \frac{1}{\tau} + \frac{T}{\beta_2} \frac{d}{d\tau}
  \left(\frac{\beta_2}{T}\right)\right]
- \frac{2}{3\beta_2 \tau} = 0~.
\ee
$\tau_{\pi}$ sets the time scale for relaxation to the {\em
  first-order} theory where $\Phi_{\rm 1st-O} = 4 \eta/3\tau$ ({\em
  not} to the ideal-fluid limit $\Phi=0$). It is related to $\eta$ and
  $\beta_2$ via $\tau_\pi = 2 \eta \beta_2$. For a classical
  Boltzmann gas of massless particles $\beta_2 = 3/(4p) = 3/(Ts)$ and so
\be \label{relaxBoltz}
\tau_\pi = \frac{6}{T} \frac{\eta}{s}~.
\ee

At infinite coupling, from the AdS/CFT correspondence $\eta/s =
1/(4\pi)$~\cite{Kovtun_1} and $\tau_\pi=(1-\log 2)/(6\pi
T)$~\cite{Heller}. For large but finite coupling, we assume that
$\eta\beta_2$ and thus $\tau_\pi$ are proportional to $\eta/s$, i.e.\
that $\beta_2=(3r)/(Ts)$ with $r=(1-\log 2)/9$; then
\be \label{relaxAdS}
\tau_\pi = r \frac{6}{T} \frac{\eta}{s}~.
\ee
Note that the numerical prefactor $r$ is about 30
times smaller than for a Boltzmann gas, implying much faster
relaxation of the dissipative fluxes to the first-order theory.

One can also define a Reynolds number via the ratio of non-dissipative
to dissipative quantities~\cite{Baym:1985tn}, $R = (e +
p)/\Phi$. Eq.~(\ref{eq:de_dtau}) can then be written as
\be\label{energy_1}
\frac{d e}{d\log \tau} = (R^{-1} - 1)(e + p) \,,
\ee
where we have neglected the bulk pressure. For stability, the
effective enthalpy $(e+p)(1-1/R)$ should be positive, i.e.\ $R>1$. The
energy density then decreases monotonically with time.

The equations of second-order dissipative fluid dynamics,
(\ref{eq:de_dtau}) or (\ref{energy_1}) and (\ref{phi}) together with
(\ref{relaxBoltz}) or (\ref{relaxAdS}) and $\beta_2 = \tau_{\pi}/(2
\eta)$ form a closed set of equations for a fluid with vanishing
currents, if augmented by an EoS.  Furthermore, the initial energy
density $e_0\equiv e(\tau_0)$ and the initial shear $\Phi_0\equiv
\Phi(\tau_0)$ have to be given. In the second-order theory one has to
specify the initial condition for the viscous stress $\Phi_0$
independently from the initial energy or particle density.  We are
presently unable to compute $\Phi_0$.  Below, we shall therefore
present results for various values of $\Phi_0$.

Alternatively, a physically motivated initial
value $\Phi_0^*$ for the stress can be obtained from the condition
that $dR/d\tau=0$ at $\tau=\tau_0$. This is the ``tipping point''
between a system that is already approaching perfect fluidity at
$\tau_0$ ($dR^{-1}/d\tau<0$ if $\Phi_0>\Phi_0^*$) and one that is
unable to compete with the expansion and is in fact departing from
equilibrium ($dR^{-1}/d\tau>0$ if $\Phi_0<\Phi_0^*$) for at least some
time after $\tau_0$.

For an EoS with constant speed of sound, say $p=e/3$, the condition
that $\dot{R}=0$ is equivalent to $\dot{e}/e = \dot{\Phi}/\Phi$.
Eqs.~(\ref{eq:de_dtau}) and~(\ref{phi}) then yield
\bea\label{ratio}
\frac{\Phi^*_0}{e_0} &=& \frac{4}{3} \frac{\tau_0}{\tau_\pi}
\left[ \sqrt{1 + \frac{4}{9r}
\frac{\tau^2_\pi}{\tau_0^2}} - 1\right] \\
&\approx& \frac{8}{27r}
\frac{\tau_\pi}{\tau_0}~~~~~~~~~~~~~~~(\tau_\pi/\tau_0\ll\sqrt{r})~,
           \label{Phi0_short_relax} \\
&=& \frac{16}{9} \frac{1}{T_0\tau_0} \frac{\eta}{s}
	   ~~~~~~~~~~~(\mbox{1st-order theory})~.\label{1stO}
\eea
The second line applies in the limit of short relaxation time; since
$\tau_\pi$ is proportional to $r$, this is always satisfied in the
limit $r\to0$. For typical initial conditions relevant for heavy-ion
collisions, it is a reasonable approximation even in the Boltzmann
limit ($r=1$). In~(\ref{1stO}) we have indicated
that~(\ref{Phi0_short_relax}) is in fact nothing but the stress in the
first-order theory (divided by the initial energy density). While it
is clear that $\Phi$ relaxes to $\Phi_{\rm 1st-O}$ over time-scales on
the order of $\tau_\pi$, eq.~(\ref{1stO}) is actually a statement
about the {\em initial} value of $\Phi$: in a fluid with reasonably
short relaxation time and stationary initial Reynolds number,
$\dot{R}(\tau_0)=0$, even the initial value of the stress is given by
the first-order approach.

The condition $R(\tau_0)>1$ for applicability of hydrodynamics
together with eq.~(\ref{Phi0_short_relax}) then provides the following
lower bound on $\tau_0$:
\be
\tau_0 > \frac{4}{3T_0} \frac{\eta}{s}\equiv \Gamma_s(\tau_0)~,
  \label{tau0_absmin}
\ee
where $\Gamma_s$ denotes the sound attenuation length; within the
first-order approach, $R=\tau/\Gamma_s$. The factor of
${\eta}/{s}$ on the right-hand-side illustrates the extended range of
applicability of hydrodynamics as compared to a Boltzmann equation: a
classical Boltzmann description requires that the thermal de-Broglie
wave length, $\sim1/T$, is smaller than the (longitudinal) size of the
system, $\tau$. For very small viscosity, though, hydrodynamics is
applicable (since $R\gg1$) even when $\Gamma_s\ll \tau \simle 1/T$.
In this point we differ somewhat from Lublinsky and Shuryak~\cite{LS07}, who
argue that the theory needs to be resummed to all orders in the
gradients of the velocity field already when $\tau\sim1/T$. From our
argument above, this should be necessary only when $\tau_0\sim\Gamma_s
(\tau_0)$, which is much smaller than $1/T_0$ if $\eta/s\ll1$.

The purpose of this paper is to motivate, however, that a much
stronger constraint than $\tau_0 > \Gamma_s(\tau_0)$ may anyhow result
from a bound on entropy production (which follows from the centrality
dependence of the multiplicity), cf.\ section~\ref{Results}.

In the Bjorken model, the entropy per unit rapidity and transverse
area at time $\tau$ is given by
\be
\frac{1}{A_\perp} \frac{dS(\tau)}{d\eta} = \tau \, \tilde{s}(\tau)~,
\ee
where $A_\perp$ is the transverse area while $\tilde s\equiv
S^{\mu}u_{\mu}$ denotes the longitudinal projection of the entropy
current. Neglecting heat flow ($q^\mu=0$) and bulk pressure ($\Pi=0$)
one obtains from~(\ref{entropy_2}):
\be
\tilde s = s \left( 1 - \frac{3}{4}\frac{\beta_2}{Ts} \, \Phi^2\right)
~. \label{entro}
\ee
$\tilde s$ can be determined, for any $\tau\ge\tau_0$, from the
solution of eqs.~(\ref{eq:de_dtau},\ref{phi}). Note that the second
term in~(\ref{entro}) is of order $(\Phi/e)^2$. For nearly perfect
fluids with $\eta/s\ll1$ and $\dot{R}(\tau_0)=0$ it is rather small.

%%%%%%%%%%%%%%%%%%%%%%%%%%%%%%%%%%%%%%%%%%%%%%%%%%%%%%%%%%%%%%%%%%%%%%%%%%%%%
\section{The CGC initial condition} \label{CGC_IC}
%%%%%%%%%%%%%%%%%%%%%%%%%%%%%%%%%%%%%%%%%%%%%%%%%%%%%%%%%%%%%%%%%%%%%%%%%%%%%

Before we can present solutions of the hydrodynamic equations, we need
to determine suitable initial conditions. To date, the most successful
description of the centrality dependence of the multiplicity is
provided by the Kharzeev-Levin-Nardi (KLN) $k_\perp$-factorization
approach~\cite{KLN}. The KLN {\em ansatz} for the unintegrated gluon
distribution functions (uGDF) of the colliding nuclei incorporates
perturbative gluon saturation at high energies and determines the
$p_\perp$-integrated multiplicity from weak-coupling QCD without
additional models for soft particle production.

Specifically, the number of gluons that are released from the
wavefunctions of the colliding nuclei is given by
\begin{eqnarray}
  \frac{dN_g}{d^2 r_{\perp}dy}&=&  {\cal N}
  \frac{N_c}{N_c^2-1} \int \frac{d^2p_\perp}{p^2_\perp}
  \int^{p_\perp} {d^2 k_\perp} \;\alpha_s(k_\perp)  \nonumber\\
  & & \times   \phi_A(x_1, (\bm{p}_\perp+\bm{k}_\perp)^2/4;
  \bm{r}_\perp)\;  \nonumber\\
  & & \times \phi_B(x_2, (\bm{p}_\perp{-}\bm{k}_\perp)^2/4; \bm{r}_\perp)~,
 \label{eq:ktfac}
\end{eqnarray}
where $N_c = 3$ is the number of colors, and $p_{\perp}$, $y$ are the
transverse momentum and the rapidity of the produced gluons,
respectively. $x_{1,2} = p_{\perp} \exp(\pm y)/\sqrt{s_{NN}}$ denote
the light-cone momentum fractions of the colliding gluon ladders, and
$\sqrt{s_{NN}}=200$~GeV is the collision energy.

The normalization factor ${\cal N}$ can be fixed from peripheral
collisions, where final-state interactions should be
suppressed. (Ideally, the normalization could be fixed from $p+p$
collisions; however, this is possible only at sufficiently high
energies, when the proton saturation scale is at least a few times
$\Lambda_{\rm QCD}$.) ${\cal N}$ also absorbs NLO corrections; when we
compare to measured multiplicities of charged hadrons, it includes as
well a factor for the average charged hadron multiplicity per gluon,
and a Jacobian for the conversion from rapidity to pseudo-rapidity.

The uGDFs are written as follows~\cite{KLN,Adil,Nara_1}:
\be  \label{uGDF}
\phi (x,k^2_{\perp}; \vec{r}_{\perp}) =
\frac{1}{\alpha_s (Q^2_s)} \frac{Q^2_s}{\textrm{max}(Q^2_s,k^2_{\perp})}
\,P(\bm{r}_{\perp})(1-x)^4~.
\ee
$P(\bm{r}_{\perp})$ denotes the probability of finding at least one
nucleon at $\bm{r}_{\perp}$~\cite{Adil,Nara_1}. This factor arises
because configurations without a nucleon at $\bm{r}_{\perp}$ do not
contribute to particle production. Note that the perturbative
$\sim1/k_\perp^2$ growth of the gluon density towards small transverse
momentum saturates at $k_\perp=Q_s$. Therefore, the
$p_\perp$-integrated gluon multiplicity obtained from~(\ref{eq:ktfac})
is finite.

We should emphasize that the {\em ansatz}~(\ref{uGDF}) is too simple
for an accurate description of high-$p_\perp$ particle production. For
example, it does not incorporate the so-called ``extended geometric
scaling'' regime above $Q_s$, which plays an important role in our
understanding of the evolution of high-$p_\perp$ spectra from
mid- to forward rapidity in d+Au collisions~\cite{DHJ}. However,
high-$p_\perp$ particles contribute little to the total multiplicity,
and more sophisticated models for the uGDF do not change the
centrality dependence of $dN/dy$ significantly~\cite{Adil}.

$Q_s(x,\bm{r}_\perp)$ denotes the saturation momentum at a given
momentum fraction $x$ and transverse coordinate $\bm{r}_{\perp}$.
It is parameterized as~\cite{Adil,Nara_1}
\begin{equation}
  Q^2_{s}(x,\bm{r}_\perp) =
  2\,{\rm GeV}^2\left(\frac{T(\bm{r}_\perp)/P(\bm{r}_\perp)}{1.53}\right)
  \left(\frac{0.01}{x}\right)^\lambda~.
  \label{eq:qs}
\end{equation}
The $\sim1/x^\lambda$ growth at small $x$ is expected from BFKL
evolution and has been verified both in deep inelastic scattering at
HERA~\cite{GBW} and in high-$p_\perp$ particle production from $d+Au$
collisions at RHIC~\cite{DHJ}; the growth speed is approximately
$\lambda\simeq0.28$. Note that the saturation momentum, as defined
in~(\ref{eq:qs}), is ``universal'' in that it doesn't depend on the
thickness of the collision partner at $\bm{r}_\perp$~\cite{LappiVenu}.

The centrality dependence of $Q_s$ is determined by the thickness
function $T(\bm{r}_\perp)$, which is simply the density distribution
of a nucleus, integrated over the longitudinal coordinate $z$. Note
that the standard Woods-Saxon density distribution is averaged over
{\em all} nucleon configurations, including those without any nucleon
at $\bm{r}_\perp$. For this reason, a factor of $1/P(\bm{r}_\perp)$
arises in $Q_s^2$~\cite{Adil,Nara_1}. It prevents $Q_s$ from dropping
to arbitrarily small values at the surface of a nucleus (since at
least one nucleon must be present at $\bm{r}_\perp$ or else no gluon
is produced at that point). The fact that $Q_s$ is bound from below
prevents infrared sensitive contributions from the surface of the
nucleus and also makes the uGDF~(\ref{uGDF}) less dependent on
``freezing'' of the one-loop running coupling.

\begin{figure}[hbt!]
\centering
\includegraphics[width=8.5cm, height = 5.9cm]{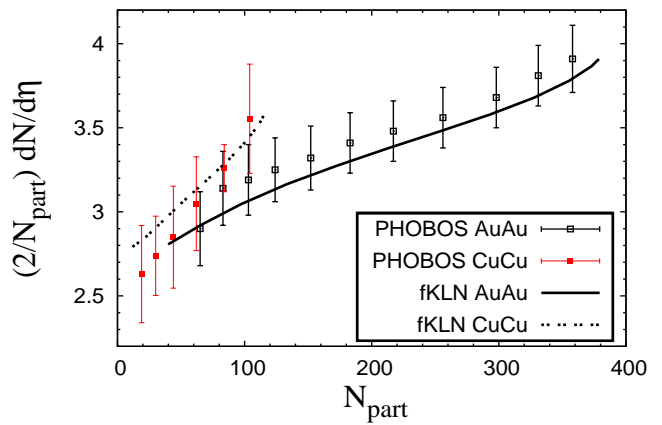}
\caption{(Color online) Centrality dependence of the charged particle
  multiplicity at midrapidity from the $k_\perp$-factorization
  approach with perturbative gluon saturation at small-$x$, for Cu+Cu
  and Au+Au collisions at full RHIC energy, $\sqrt{s_{NN}}=200$~GeV.
  PHOBOS data from ref.~\cite{phobos}; the errors are systematic, not
  statistical.}
\label{fig_1}
\end{figure}
Fig.~\ref{fig_1} shows the centrality dependence of the
multiplicity, as obtained from eq.~(\ref{eq:ktfac}) via an integration
over the transverse plane. It was noted in~\cite{Adil} that the
multiplicity in the most central collisions is significantly closer to
the data than the original KLN prediction~\cite{KLN} if the
integration over $\bm{r}_\perp$ is performed explicitly, rather than
employing a mean-field like approximation, $Q_s^2(\bm {r}_\perp)\to
\langle Q_s^2\rangle(b)$. An even better description of the data can
be obtained when event-by-event fluctuations of the positions of the
nucleons are taken into account~\cite{Nara_1}; they lead to a slightly
steeper centrality dependence of the multiplicity per participant for
very peripheral collisions or small nuclei. However, we focus here on
central Au+Au collisions and hence we neglect this effect.

It is clear from the figure that the above CGC-$k_\perp$-factorization
approach does not leave a lot of room for an additional {\em
centrality-dependent} contribution to the particle
multiplicity. (Centrality {\em independent} gluon multiplication
processes have been absorbed into ${\cal N}$.)  In fact, within the
bottom-up thermalization scenario one does expect, parametrically,
that gluon splittings increase the multiplicity by a factor $\sim
1/\alpha^{2/5}$~\cite{bottomup2} before the system thermalizes at
$\tau_0$ and the hydrodynamic evolution begins. If the scale for
running of the coupling is set by $Q_s$, this would lead to an
increase of the multiplicity for the most central Au+Au collisions by
roughly 20\%. However, such a contribution does not seem to be visible
in the RHIC data, perhaps because the bottom-up scenario, which
considered asymptotic energies, does not apply quantitatively at RHIC
energy. It is also thinkable that the model~(\ref{uGDF},\ref{eq:qs})
overpredicts the growth of the particle multiplicity per participant
with centrality somewhat.

It is noteworthy that from the most peripheral Cu+Cu to the most
central Au+Au bin, that $(dN/d\eta)/N_{\rm part}$ grows by only $\simeq
50\%$ while $N^{1/3}_{\rm part}$ increases by a factor of 2.6. Clearly,
any particle production model that includes a substantial contribution
from perturbative QCD processes will cover most of the growth. This
implies that rather little entropy production appears to occur after the
initial radiation field decoheres. If so, this allows us to correlate
the thermalization time $\tau_0$ and the viscosity to entropy density
ratio $\eta/s$. We shall assume that about 10\% entropy production may
be allowed for central Au+Au collisions.

The density of gluons at $\tau_s=1/Q_s$ is given by $dN/d^2r_\perp dy$
from eq.~(\ref{eq:ktfac}), divided by $\tau_s$. For a central
collision of Au nuclei at full RHIC energy, the average
$Q_s\simeq1.4$~GeV at midrapidity; hence $\tau_s\simeq 0.14$~fm/c. The
parton density at this time is approximately $\simeq 40$~fm$^{-3}$.
If their number is effectively conserved\footnote{We repeat that {\em
centrality-independent} gluon multiplication processes and the
contribution from quarks are already accounted for via the
normalization factor ${\cal N}$.} until thermalization at
$\tau_0$ then
\be
n(\tau_0) = \frac{\tau_s}{\tau_0}\,n(\tau_s)~.
\ee
The initial energy density $e(\tau_0)$ can now be obtained from the
density via standard thermodynamic relations. We assume that the
energy density corresponds to 16 gluons and 3 massless quark flavors
in chemical equilibrium,
\be
e(T) = \frac{47.5}{30} \pi^2 T^4~~~~,~~~~n(T)=\frac{43\zeta(3)}{\pi^2} T^3~.
\ee
\begin{figure}[hbt!]
\centering
\includegraphics[width=8.5cm, height = 5.9cm]{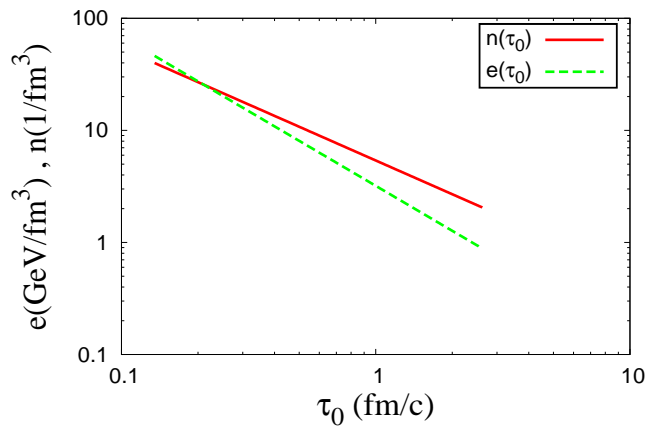}
\caption{(Color online) The parton number and energy densities
(averaged over the transverse plane) for $b=0$ Au+Au collisions at
$\sqrt{s_{NN}}=200$~GeV as functions of the thermalization time $\tau_0$.}
\label{fig_2}
\end{figure}
Fig.~\ref{fig_2} shows the parton number and energy densities at
$\tau_0$.

%%%%%%%%%%%%%%%%%%%%%%%%%%%%%%%%%%%%%%%%%%%%%%%%%%%%%%%%%%%%%%%%%%%%%%%%%%%%%
\section{Results} \label{Results}
%%%%%%%%%%%%%%%%%%%%%%%%%%%%%%%%%%%%%%%%%%%%%%%%%%%%%%%%%%%%%%%%%%%%%%%%%%%%%

%%%%%%%%%%%%%%%%%%%%%%%%%%%%%%%%%%%%%%%%%%%%%%%%%%%%%%%%%%%%%%%%%%%%%%%%%%%%%
\subsection{Evolution of the entropy and of the Reynolds number}
%%%%%%%%%%%%%%%%%%%%%%%%%%%%%%%%%%%%%%%%%%%%%%%%%%%%%%%%%%%%%%%%%%%%%%%%%%%%%

We begin by illustrating entropy production due to dissipative
effects.  Given an initial time $\tau_0$ for the hydrodynamic
evolution, we determine $\Delta S = dS_{\rm fin}/dy - dS_{\rm ini}/dy$
for $\tau> \tau_0$. This quantity increases rather rapidly at first,
since the expansion rate $H\equiv\partial_\mu u^\mu=1/\tau$ is biggest
at small $\tau$. We chose $\tau_{\rm fin}=5$~fm/c to be on the order
of the radius of the collision zone and fix the final value for
$\Delta S/S_{\rm ini}$ to equal 10\%. Having fixed the initial and
final entropy as well as the initial time then determines $\eta/s$.

\begin{figure}[hbt!]
\centering
\includegraphics[width=8.5cm, height = 5.9cm]{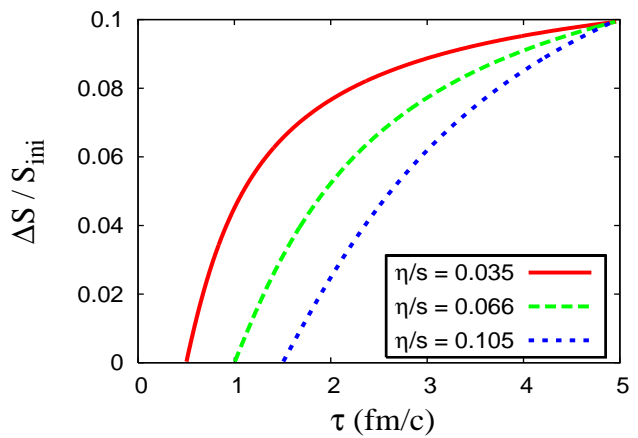}
\caption{(Color online) Entropy production within second-order dissipative
hydrodynamics as a function of (proper) time; the initial value of the
stress is $\Phi(\tau_0)=\Phi_0^*$, cf.\ eq.~(\ref{ratio}). Curves for
three different initial times, $\tau_0 = 0.5$, 1, 1.5~fm/c are shown,
and for each curve $\eta/s$ is chosen such that
$\Delta S / S_{\rm ini}=10\%$ at $\tau_{\rm fin}=5$~fm/c.}
\label{fig_3}
\end{figure}

The result of the calculation is shown in Fig.~\ref{fig_3}. As
expected, if the hydrodynamic expansion starts later (larger $\tau_0$)
then less entropy is produced for a given value of $\eta/s$;
conversely, for fixed entropy increase, larger values of $\eta/s$ are
possible. This is due to two reasons: the total time interval for
one-dimensional hydrodynamic expansion as well as the entropy
production rate decrease. In fact, the figure shows that for very
small initial time the $\Delta S/S_{\rm ini} = 10\%$ bound can not be
satisfied with $\eta/s\ge 1/(4\pi) \simeq 0.08$.

\begin{figure}[hbt!]
\centering
\includegraphics[width=8.5cm, height = 5.9cm]{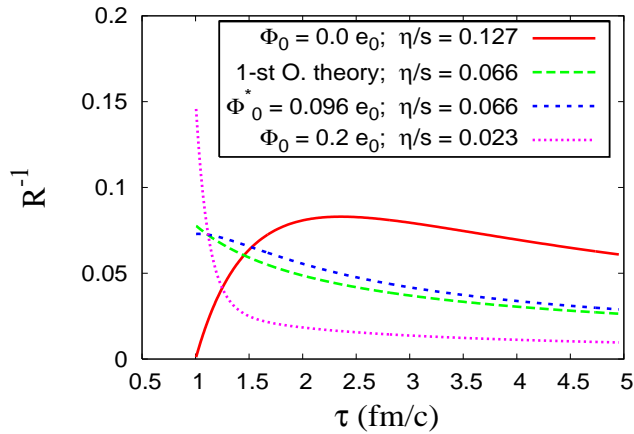}
\caption{(Color online) Time evolution of the inverse Reynolds number
  (using the Boltzmann relaxation time) for different initial values
  of the viscous stress, $\Phi_0$, and of the viscosity to entropy
  density ratio $\eta/s$. The initial time is $\tau_0=1$~fm/c for all
  curves. The short-dashed line corresponds to the initial condition
  $\dot {R}(\tau_0) = 0$, cf.\ eq.~(\ref{ratio}). The long-dashed line
  corresponds to the first-order theory.}
\label{fig_4}
\end{figure}
In Fig.~\ref{fig_4} we show the behavior of the inverse Reynolds
number for different initial values of the stress. Again, for each
curve $\eta/s$ is fixed such that $\Delta S/S_{\rm ini} = 10\%$ at
$\tau_{\rm fin}=5$~fm/c. As already indicated above, if
$\Phi_0<\Phi_0^*$ defined in eq.~(\ref{ratio}), the fluid can not
compete with the expansion and departs from equilibrium. On the other
hand, if $\Phi_0>\Phi_0^*$, there is already a rapid approach towards
the perfect-fluid limit at $\tau_0$. In either case, the
interpretation of $\tau_0$ as the earliest possible starting time for
hydrodynamic evolution does not appear sensible. The initial
condition corresponding to $\dot{R}(\tau_0)=0$ in turn corresponds to
the situation where the fluid has just reached the ability to approach
equilibrium. It is clear from the figure that the evolution is close
to that predicted by the first-order theory.

\begin{figure}[hbt!]
\centering
\includegraphics[width=8.5cm, height = 5.9cm]{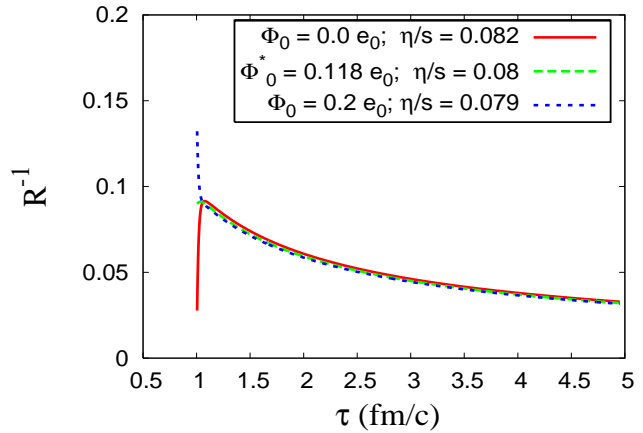}
\caption{(Color online) Same as Fig.~\ref{fig_4} but for the
  generalized AdS/CFT relaxation time~(\ref{relaxAdS}).}
\label{fig_5}
\end{figure}
Fig.~\ref{fig_5} shows the Reynolds number for our {\em
  ansatz}~(\ref{relaxAdS}) for the relaxation time at strong coupling,
  which essentially follows the behavior given by the first-order
  theory: after a time $\sim\tau_\pi$ has elapsed, the behavior of $R$
  is nearly independent of the initial value of $\Phi$. The initial
  condition $\Phi_0=\Phi_0^*$ again leads to the most natural behavior
  of $R$ without a very rapid initial evolution.

%%%%%%%%%%%%%%%%%%%%%%%%%%%%%%%%%%%%%%%%%%%%%%%%%%%%%%%%%%%%%%%%%%%%%%%%%%%%%%
\subsection{$\eta/s$ versus $\tau_0$}
%%%%%%%%%%%%%%%%%%%%%%%%%%%%%%%%%%%%%%%%%%%%%%%%%%%%%%%%%%%%%%%%%%%%%%%%%%%%%%

From the previous results it is evident that fixing the amount of
produced entropy, $\Delta S/S_{\rm ini}$, correlates $\eta/s$ with
$\tau_0$. In this section we show the upper limit of $\eta/s$ as a
function of $\tau_0$.

\begin{figure}[hbt!]
\centering
\includegraphics[width=8.5cm, height = 5.9cm]{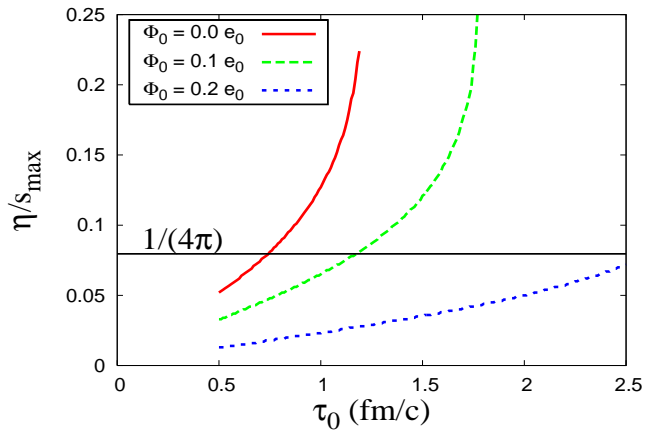}
\caption{(Color online) Bound on $\eta/s$ as a function of $\tau_0$,
for $\Delta S/S_{\rm ini} = 10\% $ entropy production, for a Boltzmann
gas.}
\label{fig_6}
\end{figure}
We begin with the Boltzmann gas with fixed $\Phi_0$ (independent of
$\tau_0$) in Fig.~\ref{fig_6}. One observes that the maximal viscosity
depends rather strongly on the initial value of the stress. For any
given $\Phi_0$, $(\eta/s)_{\rm max}$ first grows approximately
linearly with $\tau_0$. For large initial time, however, the expansion
and entropy production rates drop so much that the bound on viscosity
eventually disappears. Furthermore, it is interesting to observe that the
conjectured lower bound $\eta/s=1/(4\pi)$ excludes too rapid
thermalization: even if the fluid is initially perfectly equilibrated
($\Phi_0=0$), a thermalization time well below $\sim1$~fm/c is
possible only if either $\eta/s<1/(4\pi)$ or $\Delta S/S_{\rm
ini}>10\%$. With $10\%$ corrections to perfect fluidity at $\tau_0$,
shown by the long-dashed line in Fig.~\ref{fig_6}, the minimal
$\tau_0$ compatible with both $\eta/s\ge1/(4\pi)$ and $\Delta S/S_{\rm
ini}=10\%$ is about 1.2~fm/c.  If $\eta/s \simeq 0.1-0.2$, as deduced
from the centrality dependence of elliptic flow at RHIC~\cite{DDGO},
then $\tau_0\simeq1.5$~fm/c.

\begin{figure}[hbt!]
\centering
\includegraphics[width=8.5cm, height = 5.9cm]{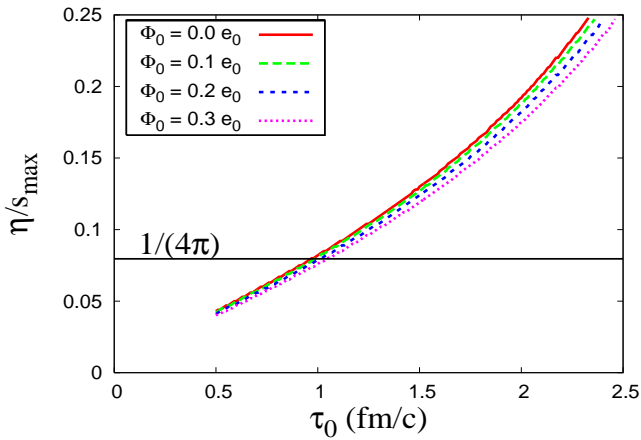}
\caption{(Color online) Same as Fig.~\ref{fig_6} but for strong
  coupling, $\tau_\pi$ from~(\ref{relaxAdS}).}
\label{fig_7}
\end{figure}
In Fig.~\ref{fig_7} we perform a similar analysis for our {\em
ansatz}~(\ref{relaxAdS}) for the strong-coupling case. Due to the much
smaller relaxation time of the viscous stress, we observe that the
viscosity bound is now rather insensitive to the magnitude of the
initial correction to equilibrium.  We obtain a lower bound on the
thermalization time of $\tau_0 \simeq 1$~fm/c for the minimal
viscosity, increasing to about 1.2 - 2.2~fm/c if $\eta/s\simeq$ 0.1 -
0.2.

\begin{figure}[hbt!]
\centering
\includegraphics[width=8.5cm, height = 5.9cm]{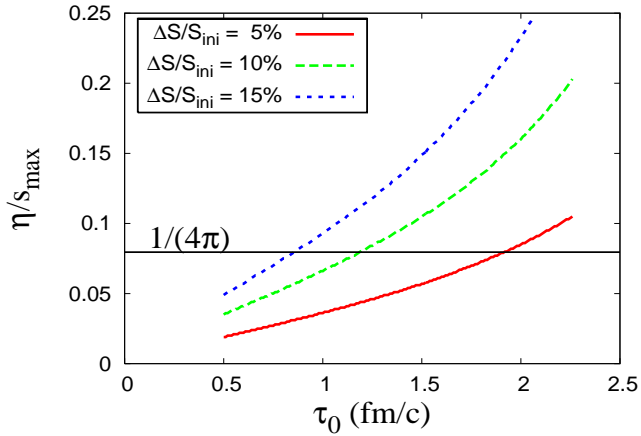}
\caption{(Color online) Same as Fig.~\ref{fig_6} but for
  $\Phi_0=\Phi_0^*$ and different bounds on entropy production.}
\label{fig_8}
\end{figure}
In Fig.~\ref{fig_8} we return to the Boltzmann gas with initial value
for the stress as given in eq.~(\ref{ratio}), corresponding to
$\dot{R}(\tau_0) =0$. Comparing to Fig.~\ref{fig_6}, we observe that
the viscosity bound is affected mostly for large $\tau_0$: with this
initial condition, a high viscosity $\eta/s\sim1$ is excluded even if
the initial time is as big as 2~fm/c. The reason why the upper bound
on the viscosity does not disappear at large $\tau_0$ for this initial
stress is that $\Phi_0^*/e_0$ grows with $\eta/s$, cf.\
eq.~(\ref{ratio}). A lot of entropy would then be produced, even for
large $\tau_0$.

\begin{figure}[hbt!]
\centering
\includegraphics[width=8.5cm, height = 5.9cm]{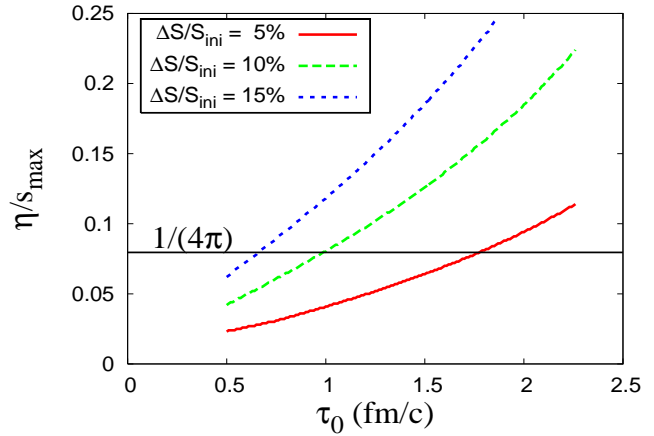}
\caption{(Color online) Same as Fig.~\ref{fig_8} but for the
  strong-coupling limit, i.e.\ for $\tau_\pi$ from eq.~(\ref{relaxAdS}).}
\label{fig_9}
\end{figure}
Fig.~\ref{fig_8} also gives an impression of the sensitivity to the
entropy production bound. For $\eta/s=0.15$, for example, $\tau_0$
decreases from $\simeq1.8$~fm/c, if the entropy is allowed to increase
by 10\%, to $\simeq1.5$~fm/c if the bound is relaxed to $\Delta
S/S_{\rm ini}=15\%$.

We performed similar calculations for the strong-coupling limit as
shown in Fig.~\ref{fig_9}. The curves are rather close to those for a
Boltzmann gas from Fig.~\ref{fig_8}, which is expected. With this
initial condition, i.e.\ $\Phi_0=\Phi_0^*$, the hydrodynamic evolution
is close to the first-order theory for both cases. Entropy production
is sensitive only to $\tau_0$ and $\eta/s$ but is nearly independent
of the stress relaxation time $\tau_\pi$.

\begin{figure}[hbt!]
\centering
\includegraphics[width=8.5cm, height = 5.9cm]{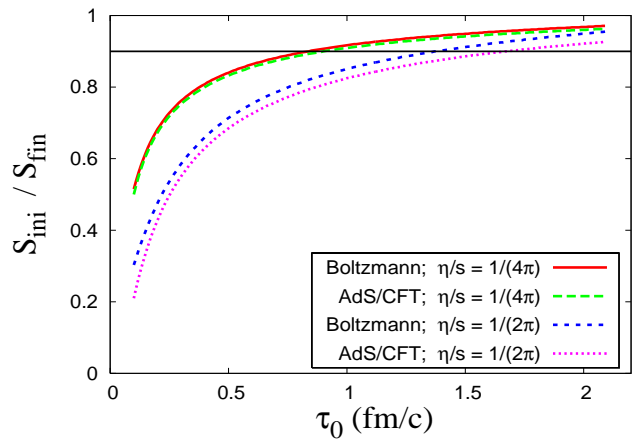}
\caption{(Color online) Initial over final entropy as a function of
  $\tau_0$ for two different values of $\eta/s$, and two different
  relaxation times corresponding to weak and strong coupling.}
\label{fig_10}
\end{figure}
Fig.~\ref{fig_10} shows the ratio of the final to the initial entropy
as a function of $\tau_0$ for two different values of $\eta/s$, and
the two different relaxation times discussed above in
eqs.~(\ref{relaxBoltz}, \ref{relaxAdS}). Here, $S_{\rm fin}$ has been
fixed to the value appropriate for central Au+Au collisions while
$S_{\rm ini}$ is varied accordingly. For example,
for $\tau_0=0.6$~fm/c and $\eta/s=1/(2\pi)$, almost 30\% entropy
production occurs. This would account for the entire growth of
$(dN/d\eta)/N_{\rm part}$ from $N_{\rm part}\simeq60$ to $N_{\rm
part}\simeq360$ observed in Fig.~\ref{fig_1}. That is, for these
parameters the initial parton multiplicity per participant would have
to be completely independent of centrality. For the same viscosity,
$\tau_0=0.3$~fm/c would imply that nearly half of the final-state
entropy was produced during the hydrodynamic stage, i.e.\ that the
initial multiplicity per participant should actually {\em decrease}
with centrality. Such a scenario appears unlikely to us. Note that
even for $\tau_0=0.3$~fm/c and $\eta/s=1/(4\pi)$, with $T=400$~MeV one
finds that $\Gamma_s(\tau_0)/\tau_0\simeq0.17$ is quite small.
Romatschke obtained similar numbers for the initial to final
entropy ratio, albeit only for $\tau_0=1$~fm/c, in a computation that
included cylindrically symmetric transverse expansion~\cite{Romatschke2}.

%%%%%%%%%%%%%%%%%%%%%%%%%%%%%%%%%%%%%%%%%%%%%%%%%%%%%%%%%%%%%%%%%%%%%%%%%%%%%
\section{Summary, Discussion and Outlook} \label{Summary}
%%%%%%%%%%%%%%%%%%%%%%%%%%%%%%%%%%%%%%%%%%%%%%%%%%%%%%%%%%%%%%%%%%%%%%%%%%%%%

In this paper, we have analyzed entropy production due to non-zero shear
viscosity in central Au+Au collisions at RHIC. We point out that a
good knowledge of the {\em initial conditions}, and of the final
state, of course, can provide useful constraints for hydrodynamics of
high-energy collisions, specifically on transport coefficients, on the
equation of state (not discussed here, cf.~\cite{DDGO}), on the
initial/thermalization time and so on.

Our main results are as follows. Assuming that hydrodynamics applies
at $\tau>\Gamma_s$, then due to the rather restrictive bound on
entropy production, it follows that the shear viscosity to entropy
density ratio of the QCD matter produced at central rapidity should be
small, at most a few times the lower bound $\eta/s=1/(4\pi)$ conjectured
from the AdS/CFT correspondence at infinite coupling. This represents
a consistency-check with similar numbers ($\eta/s\simle0.2$)
extracted from azimuthally asymmetric elliptic
flow~\cite{Teaney,Lacey:2006bc,DDGO,RR07}. We have neglected several
other possible sources of entropy production, such as bulk viscosity
near the transition region~\cite{bulk} or hadronic corona
effects~\cite{HiranoGy}; such additional contributions might tighten
the constraints even further.

Furthermore, the entropy production bound correlates the maximal
allowed viscosity to the initial time $\tau_0$ for hydrodynamic
evolution. This is due to the fact that the expansion rate is equal to
the inverse of the expansion time, which makes entropy production from
viscous effects rather sensitive to the value of $\tau_0$. We have
found that for $\Delta S/S_{\rm ini}\simeq10\%$, that the initial time
for hydrodynamics should be around 1~fm/c, possibly a little
larger. Significantly smaller thermalization times would either
require $\eta/s<0.15-0.2$ (or even smaller than $1/(4\pi)$). Alternatively,
they would require a particle production mechanism that yields
significantly lower initial multiplicities than the KLN-CGC
approach. Given the very good description of the centrality dependence
of the multiplicity, however, to us it appears reasonable to assume that
this approach provides an adequate initial condition in that
the initial parton multiplicity per participant increases with centrality.

A significant problem with viscous hydrodynamics, in particular with
the second-order approach of Israel-Stewart, is the fact that the
number of initial parameters increases. Even within the most simple
framework followed here (1+1D Bjorken expansion combined with neglect
of conserved currents, of bulk viscosity, and of heat flow), a unique
solution requires us to specify, in addition to the ideal-fluid
parameters, the shear viscosity and the initial value for the stress. The
latter, in particular, is not a general property of near-equilibrium
QCD but depends on the parton liberation and thermalization
process. We have, however, introduced a physically motivated initial
condition for the stress: if $\tau_0$ is defined as the earliest possible
initial time for hydrodynamics, it is plausible that the initial
Reynolds number should be stationary, $\dot{R}(\tau_0)=0$. Otherwise the
fluid either still departs from equilibrium ($\dot{R}(\tau_0)<0$); or
is already approaching it ($\dot{R}(\tau_0)>0$).

For small relaxation times of the stress, the condition that
$\dot{R}(\tau_0)=0$ implies that its initial value already be close to
that given by the first-order theory of Eckart, Landau and Lifshitz
(the relativistic generalization of Navier-Stokes hydrodynamics). We
therefore expect that in general the two approaches will provide
rather similar results for heavy-ion collisions. One should keep in
mind, however, that in the second-order theory the entropy current
includes a term quadratic in the stress, which is of course absent
from the first-order theory, and which reduces entropy production
slightly.

Perhaps most importantly, with $\dot{R}(\tau_0)=0$, the hydrodynamic
evolution is largely independent of the stress relaxation time
$\tau_\pi$, and therefore similar for both a Boltzmann gas at weak
coupling (with low viscosity, however) and a strongly coupled plasma.
The latter relaxes very rapidly to the first-order theory, regardless
of the initial condition. The former, on the other hand, is forced by
the initial condition to start close to relativistic Navier-Stokes,
and the relaxation time is still sufficiently small to prevent a
significant departure from the first-order theory.

The initial condition $\dot{R}(\tau_0)=0$ also guarantees that
$R(\tau) \gg1$ for all $\tau\ge\tau_0$, as long as the initial time is
not extremely short ($\tau_0 T_0 \gg \eta/s$). The effective enthalpy
$(1-1/R)(e+p)$ is therefore always positive. On the other hand, our
numerical results indicate that the Reynolds number does not exceed
$\sim100$ during the QGP phase. This is well below the regime where
Navier-Stokes turbulence occurs in incompressible, non-relativistic
fluids ($R\simge 1000$). Indeed, turbulence during the hydrodynamic
stage would probably cause large fluctuations of the elliptic flow
$v_2$~\cite{TB}, which are not seen~\cite{Sorensen:2006nw}.

A quantitative interpretation of hydrodynamic flow effects in
heavy-ion collisions at RHIC and LHC will of course require 2+1D and
3+1D
solutions~\cite{RR07,Heinz,MuroRi,BRW,Romatschke,Romatschke2,Muronga2}. The
results obtained here should prove useful for constraining the initial
conditions (in particular $\tau_0$ and $\Phi_0$) for such large-scale
numerical efforts. In particular, as we pointed out here, the entropy
production bound correlates $\tau_0$ with $\eta/s$. In turn, we expect
that elliptic flow will provide an {\em anti-correlation} since later
times and larger shear viscosity should both reduce its magnitude. The
intersection of those curves could then provide an estimate of the
initial time for hydrodynamics at RHIC.

%%%%%%%%%%%%%%%%%%%%%%%%%%%%%%%%%%%%%%%%%%%%%%%%%%%%%%%%%%%%%%%%%%%%%%%%%%%%%%
\section*{ACKNOWLEDGMENTS}
The authors would like to thank L. P. Csernai, H. J. Drescher, M.~Gyulassy,
D.~H.~Rischke, D.~Schiff, and H.~St\"ocker for useful discussions.
E. M.\ gratefully acknowledges support by the Alexander von Humboldt foundation.
%%%%%%%%%%%%%%%%%%%%%%%%%%%%%%%%%%%%%%%%%%%%%%%%%%%%%%%%%%%%%%%%%%%%%%%%%%%%%

\end{document}